\documentclass[showpacs,preprintnumbers,amsmath,amssymb]{revtex4}
\begin{document}
\preprint{IMAFF-RCA-05-07}
\title{Wormholes in the accelerating universe}

\author{Pedro F. Gonz\'{a}lez-D\'{\i}az and Prado Mart\'{\i}n-Moruno}

\affiliation{Colina de los Chopos, IMAFF, CSIC,\\ Serrano 121,
28006 Madrid (SPAIN) \email{p.gonzalezdiaz@imaff.cfmac.csic.es}}

\begin{abstract}
We discuss different arguments that have been raised against the
viability of the big trip process, reaching the conclusions that
this process can actually occur by accretion of phantom energy
onto the wormholes and that it is stable and might occur in the
global context of a multiverse model. We finally argue that the
big trip does not contradict any holographic bounds on entropy and
information.
\end{abstract}


\maketitle


1. We shall consider in more detail first how the big trip
\cite{Gonzalez-Diaz:2006na} can be derived when a simple non
static Morris-Thorne metric is used for a wormhole, i. e.
\begin{equation}\label{uno}
{\rm d}s^2=-{\rm d}t^2+\frac{{\rm
d}r^2}{1-\frac{K(r,t)}{r}}+r^2\left({\rm d}\theta^2+{\rm
sin}^2\theta{\rm d}\phi^2\right),
\end{equation}
where we have taken the shift function to be zero and we let
\cite{Gonzalez-Diaz:2006na} the shape function $K$ to also depende
on time. If dark energy is regarded to be a perfect fluid with
$T_{\mu\nu}=(p+\rho)u_{\mu}u_{\nu}+pg_{\mu\nu}$, with $p$ and
$\rho$ the pressure and energy density, respectively, and
$u^{\mu}={\rm d}x^{\mu}/{\rm d} s$ is the four-velocity,
$u^{\mu}u_{\mu}=-1$, the conservation law for the time-component
of the energy-momentum tensor, $T^{\nu}_{\mu ;\nu}=0$, can be
integrated over $r$ to give
\begin{equation}\label{dos}
ur^2m^{-2}\left(1-\frac{K(r,t)}{r}\right)^{-1}\left(1-\frac{K(r,t)}{r}+u^2\right)^{1/2}(p+\rho)e^{\int^{r}_{\infty}{\rm
d}r\alpha }=C(t),
\end{equation}
in which we have introduced the exotic mass factor $m^{-2}$ to
provide the r.h.s. function $C(t)$ with the dimension of an energy
density, and
\begin{equation}\label{tres}
\alpha=\frac{\partial_0T^0_0}{T^r_0}+\frac{\partial_0K(r,t)}{2r\left(1-\frac{K(r,t)}{r}\right)}\frac{T^0_0-T^r_r}{T^r_0}.
\end{equation}
Integrating then over $r$ the conservation law for energy-momentum
tensor projected on four-velocity, $u_{\mu}T^{\mu\nu}_{;\nu}=0$,
we have
\begin{equation}\label{cuatro}
r^2u\left(1-\frac{K(r,t)}{r}\right)^{-1/2}e^{\int^{\rho}_{\rho_{\infty}}\frac{{\rm
d}\rho }{p(\rho)+\rho}}e^{\int^r_{\infty}{\rm d}r\beta}=A(t),
\end{equation}
in which $A(t)$ is a function of time having the dimension of a
squared mass satisfying that $A(t)=\lim_{r\rightarrow\infty}ru^2$
does not depend on the radial coordinate and does on t only
through the mass $m$, so that $A(t)=A'm^2$, $A'$ being a
dimensionless positive ($u>0$) constant; finally
\begin{equation}\label{cinco}
\beta
=\frac{\left(1-\frac{K(r,t)}{r}+u^2\right)^{1/2}}{u\left(1-\frac{K(r,t)}{r}\right)^{1/2}}\left[\frac{\partial_0\rho}{p+\rho}+\frac{\partial_0K(r,t)}{2r\left(1-\frac{K(r,t)}{r}\right)}\right]+\frac{\partial_0\left(\left(1-\frac{u^2}{1-\frac{K(r,t)}{r}}\right)^{1/2}\right)}{u}
\end{equation}
From Eqs.(\ref{dos}) and (\ref{cuatro}) we get
\begin{equation}\label{seis}
(p+\rho)\left(1-\frac{K(r,t)}{r}\right)^{-1/2}\left(1-\frac{K(r,t)}{r}+u^2\right)^{1/2}e^{\int^{\rho}_{\rho_{\infty}}\frac{{\rm
d}\rho }{p(\rho)+\rho}}e^{\int^r_{\infty}{\rm d}r
(\alpha-\beta)}=B(t),
\end{equation}
with $B(t)=C(t)/A'=p[\rho_{\infty}(t)]+\rho_{\infty}(t)$. The rate
of exotic mass due to phantom energy accretion should be given by
integrating over ${\rm d}S=r^2\sin\theta{\rm d}\theta{\rm d}\phi$
the nonzero component $T^r_0$, $\dot{m}=\int{\rm d}ST^r_0$, the
sign being chosen to account for accreating negative energy.
Taking into account Eqs.(\ref{cuatro}) and (\ref{seis}), we obtain
\begin{equation}\label{siete}
\dot{m}=-4\pi(p+\rho)A'm^2\left(1-\frac{K(r,t)}{r}\right)^{1/2}e^{-\int^r_{\infty}{\rm
d}r\alpha }.
\end{equation}
It follows that in the asymptotic limit $r\rightarrow\infty$, in
which the exponent in Eq. (\ref{siete}) vanishes, this rate
reduces to
\begin{equation}\label{ocho}
\dot{m}=-4\pi A'm^2(p+\rho).
\end{equation}
Inserting the energy density for a general quintessence fluid with
$p=w\rho$ \cite{Gonzalez-Diaz:2004eu}, for $w<-1$ (phantom energy)
in Eq.(\ref{ocho}) we finally derive for the time-dependent exotic
mass
\begin{equation}\label{nueve}
m=m_0\left[1-\frac{4\pi
A'(|w|-1)\rho_0m_0(t-t_0)}{1-\frac{3}{2}C(|w|-1)(t-t_0)}\right]^{-1},
\end{equation}
where the "0" subscripts mean current values and
$C=\left(8\pi\rho_0/3\right)^{1/2}$. Hence, a big trip where the
wormhole throat diverges will take place before the occurrence of
the big rip singularity, at a time
\begin{equation}\label{diez}
t_*=t_0+\frac{t_{{\rm
br}}-t_0}{1+\left(8\pi\rho_0/3\right)^{1/2}A'm_0}<t_{{\rm br}},
\end{equation}
in which $t_{{\rm br}}=t_0+\frac{2}{3(|w|-1)C}$ is the big rip
time. So, during a given time interval before $t_*$ the size of
the wormhole throat will exceed that of the universe.

Formally speaking, the above procedure does not take into account
the feature that we are not dealing with a vacuum solution, such
as Faraoni has recently pointed out \cite{Faraoni:2007kx}.
However, all our calculations are finally referred to the
asymptotic case $r\rightarrow\infty$, where the r.h.s. of
$\lambda'\left(e^{-\lambda}-1\right)/r=\Theta_{11}$, which is
obtained from the Einstein equations for an ansatz ${\rm
d}s^2=-{\rm d}t^2+e^{\lambda}{\rm d}r^2+r^2{\rm d}\Omega^2_2$,
 vanishes
because $\Theta_{11}={\rm constant}/r^4$ for solution (\ref{uno}),
where we can still keep $\Theta_{00}=0$. It follows that Eq.
(\ref{ocho}) is correct if the big trip is defined for an
asymptotic observer.

However, the most serious argument against the occurrence of the
big trip in the universe most recently raised by Faraoni
\cite{Faraoni:2007kx} is that the accretion of phantom energy with
a perfect fluid equation of state is characterized by a radial
velocity $v_{R}\sim a^{3(1+w)/2}$ which strictly vanishes at the
big rip singularity and in any event quickly decreases with time
for $w<-1$. Thus, according to Faraoni, also at the time where the
big trip would occur, accretion of phantom energy would be largely
prevented and the big trip phenomenon would not take place at all.
Besides the feature that the size of the wormhole throat equalizes
that of the Universe before it diverges, what matters here is not
the fluid velocity but its flow (as expressed as phantom energy
per unit surface per unit time) which can be roughly given by
$v_{R}\rho$, that is $\sim a^{-3(1+w)/2}$, which in fact increases
with time and consistently diverges at the big rip. Then the
argument by Faraoni does not apply to the case and the big trip
can not be dismissed due to it.

On the other hand and even more importantly, what we are dealing
here with is no longer accretion of usual energy concentrated on
given regions of space, but vaccum energy which isotropically and
homogeneously pervades the whole space, even the regions ocuppied
by physical objects. Hence, accretion of phantom energy is not
based on any fluid motion but on increasing more and more space
filled with phantom energy inside the throat. The big trip
phenomenon would then appear when one superposes to this effect
the feature that the phantom energy density increases with time.

2. Since the wormhole spacetime is asymptotically flat the big
trip process has debatably been considered to take place in the
framework of the multiverse where the mouth of a grown up wormhole
can still be inserted in larger universes.

3. Wormholes undergoing a big trip process are
quantum-mechanically stable because the parameter $\xi$
characterizing the regularized Hadamard function, $\langle
\Theta_{\mu\nu}\rangle_{{\rm reg}}\sim{\rm const}/\xi^4$ should
necessarily be nonvanishing during the process.

4. The Bekenstein bound on information and entropy could pose a
further problem if the final time for the phantom universe is
taken to be that for the big rip. However, in the neighborhood of
the big rip, small wormholes would crop up and be connected to the
region after the big rip in such a way that any amount of
information is actually allowed to be transferred in the big trip.

The big trip process is a rather weird phenomenon which shows some
paradoxical consequences. Actually, one could in principle expect
such consequences and even the big trip itself to be avoided by a
proper quantum gravity treatment, as that process takes place as
one is approaching the big rip singularity where the energy
density becomes infinite. However, since we have not still a
proper quantum theory of gravitation nothing can said for sure
concerning that possibility.

\vfill

\end{document}